\documentclass[fleqn,11pt]{article}

\usepackage{amsmath,amssymb,amsfonts,cite}

\def\noi{\noindent}

\newcommand{\Title}[1]{\noi {{\Large\bf #1}}\\[1ex]}

\def\Aunames#1{\noi{\bf #1}}
\def\auth#1{${}^{#1}$}
\def\Addresses#1{\medskip\noi \protect
	\begin{description}\itemsep -3pt {\it #1} \end{description}}
\def\addr#1#2{\item[${}^{#1}$]{\it #2}}

\newcommand{\Abstract}[1]{\vskip 2mm \begin{center}
        \parbox{16.4cm}{\small\noi #1} \end{center}\medskip}

\def\nqq{\hspace*{-2em}}

\def\cm{\hspace*{1cm}}
\def\inch{\hspace*{1in}}

\def\Jl#1#2{#1 {\bf #2},\ }

\def\ApJ#1 {\Jl{Astroph. J.}{#1}}
\def\CQG#1 {\Jl{Class. Quantum Grav.}{#1}}
\def\DAN#1 {\Jl{Dokl. AN SSSR}{#1}}
\def\GC#1 {\Jl{Grav. Cosmol.}{#1}}
\def\GRG#1 {\Jl{Gen. Rel. Grav.}{#1}}
\def\JETF#1 {\Jl{Zh. Eksp. Teor. Fiz.}{#1}}
\def\JETP#1 {\Jl{Sov. Phys. JETP}{#1}}
\def\JHEP#1 {\Jl{JHEP}{#1}}
\def\JMP#1 {\Jl{J. Math. Phys.}{#1}}
\def\NPB#1 {\Jl{Nucl. Phys. B}{#1}}
\def\NP#1 {\Jl{Nucl. Phys.}{#1}}
\def\PLA#1 {\Jl{Phys. Lett. A}{#1}}
\def\PLB#1 {\Jl{Phys. Lett. B}{#1}}
\def\PRD#1 {\Jl{Phys. Rev. D}{#1}}
\def\PRL#1 {\Jl{Phys. Rev. Lett.}{#1}}

\def\lal{&&\nqq {}}
\def\eq{Eq.\,}
\def\eqs{Eqs.\,}
\def\beq{\begin{equation}}
\def\eeq{\end{equation}}
\def\bear{\begin{eqnarray}}
\def\bearr{\begin{eqnarray} \lal}
\def\ear{\end{eqnarray}}
\def\earn{\nonumber \end{eqnarray}}

\def\nnn{\nonumber\\ \lal }

\def\yyy{\\[5pt] \lal }

\def\sequ#1{\setcounter{equation}{#1}}

\def\dst{\displaystyle}
\def\tst{\textstyle}
\def\fracd#1#2{{\dst\frac{#1}{#2}}}
\def\fract#1#2{{\tst\frac{#1}{#2}}}

\def\half{{\fract{1}{2}}}

\def\e{{\,\rm e}}
\def\d{\partial}

\def\diag{\mathop{\rm diag}\nolimits}

\def\const{{\rm const}}
\def\eps{\varepsilon}

\def\R{{\mathbb R}}
\def\mn{_{\mu\nu}}
\def\mN{_{\mu}^{\nu}}

\def\sph{spherically symmetric}
\def\ssph{static, spherically symmetric}
\def\wh{wormhole}
\def\whs{wormholes}
\def\RN{Reissner--Nordsrtr\"om}
\def\GR{general relativity}

\begin{document}

\twocolumn[
%\jnumber{4}{2013}

\Title{Example of a stable wormhole in general relativity}

\Aunames{K. A. Bronnikov,\auth{a,b,1} L. N. Lipatova,\auth{c}
         I. D. Novikov,\auth{c,d} and A. A. Shatskiy\auth{c}}

\Addresses{ \small
\addr a {Center for Gravitation and Fundamental Metrology, VNIIMS,
             Ozyornaya ul. 46., Moscow 119361, Russia}
\addr b {Institute of Gravitation and Cosmology, PFUR,
             ul. Miklukho-Maklaya 6, Moscow 117198, Russia}
\addr c {Astro Space Center, Lebedev Physical Institute of RAS,
         Profsoyuznaya ul. 84/32, Moscow 117997, Russia}
\addr d {Niels Bohr International Academy, Niels Bohr Institute,
         Blegdamsvej 17, DK-2100 Copenhagen, Denmark}
         }

\Abstract
  {We study a \ssph\ \wh\ model whose metric coincides with that of the
  so-called Ellis \wh\ but the material source of gravity consists of a
  perfect fluid with negative density and a source-free radial electric or
  magnetic field. For a certain class of fluid equations of state, it has
  been shown that this \wh\ model is linearly stable under both \sph\
  perturbations and axial perturbations of arbitrary multipolarity.
  A similar behavior is predicted for polar nonspherical perturbations.
  It thus seems to be the first example of a stable \wh\ model in the
  framework of general relativity (at least without invoking phantom thin
  shells as \wh\ sources).}

] %%%%%%%%%%%%%%%%%%%%%%%%%%%%%%%%%%%%%%%%%%%

\section{Introduction}
\label{br-s1}

  The stability of any static or stationary object under small perturbations
  is a necessary condition for its steady existence in the Universe.
  Traversable Lorentzian \whs, being a subject of substantial attention in
  the modern research in \GR\ and its extensions, are not an exception, and
  much effort has been applied to their stability studies, see, e.g., [1--10].
%   \cite{Armendariz-Picon2002, Shinkai2002, Gonzalez2008-1,
%   Gonzalez2008-2, Gonz-09, Gri-01, Gri-02,
%   Doroshkevich2009-1, Novikov2009-2, Sarbach2010}.
  Stable \wh\ models have been obtained in some generalized theories
  of gravity (see, e.g., [11]), but, to our knowledge, in \GR\ such
  examples have not been found by now (at least without invoking phantom
  thin shells as \wh\ sources).

  The simplest (zero-mass) \wh\ model with the metric
\bearr
    ds^2 = dt^2-dx^2-r^2(x)[d\theta^2 + \sin^2\theta\,d\varphi^2] ,
\nnn
    \quad r^2(x) = q^2+x^2                              \label{br-ds0}
\end{eqnarray}
  and a material source in the form of a massless scalar field with negative
  energy (a phantom scalar field), obtained in [12, 13]
%  \cite{Bronnikov1973, Ellis1973}
  and also discussed by Morris and Thorne [14], %%\cite{Morris1988-2},
  turned out to be unstable, contrary to a conclusion of [1]
%  \cite{Armendariz-Picon2002}
  (where not all possible perturbations were
  considered) and according to later, more complete studies [2--4, 8, 15]
%  \cite{Shinkai2002, Gonzalez2008-1, Gonzalez2008-2, Doroshkevich2009-1,
%  Bronnikov2011},
  which also allowed for nonzero \wh\ masses. As follows from [5], inclusion
  of an electric or magnetic charge (the corresponding exact \wh\ solutions
  in \GR\ and scalar-tensor gravity are known from [12]) does not stabilize
  \whs\ supported by a phantom scalar.

  Evidently, the stability properties of different configurations
  with the same metric but with other kinds of matter should also be
  different, depending on the particular dynamics of the material source.
  Accordingly, a \wh\ with the same metric (\ref{br-ds0}) but another material
  source, a radial magnetic field and phantom dust with negative mass
  density, was studied in [8],
% \cite{Doroshkevich2009-1}
  and it turned out to be stable under all spherical perturbations except for
  inertial radial motion of dust particles. The unstable mode grows
  slowly enough, linearly in time [9]. It was also shown [10] that nonlinear
  perturbations of this model lead to shell-crossing singularities.

  It was later conjectured that the instabilities of this model can be damped
  by introducing an additional parameter, related to a nonzero pressure
  proportional to a deflection from the background static configuration.
  This created a hope to construct a completely stable model in which the
  unstable mode would be absent and where any shell crossing would be
  prevented by repulsive hydrodynamic forces.

  Following this idea, in this paper we study the stability of a \wh\ model
  with the metric (\ref{br-ds0}) and a matter source in the form of a radial
  monopole magnetic (or electric) field and a perfect phantom fluid whose
  equation of state is close to that of dust. The electromagnetic field has
  no source (``a charge without charge'' according to Wheeler [16]), but its
  lines of force extend from one spatial infinity to the other.\footnote
  	{Similar configurations with phantom dust were previously
	considered as possible nonsingular classical particle models [17].}
  In the static configuration, whose stability is under study, the fluid
  energy density $\eps$ is negative, and its absolute value is twice the
  energy density of the magnetic field. The pressure $p$ is absent in the
  static case (phantom dust) but grows proportionally to the difference of
  the perturbed density from its static value.

  A tentative study, indicating the stability of such a \wh\ under spherical
  perturbations, was performed in [18], but the perturbation mode with a
  changing throat radius was not considered there. In the present paper we
  will prove the stability of a class of such models under all spherical and
  all axial nonsherical perturbations. We will also speculate on the
  behavior of polar nonspherical perturbations and conclude that they should
  also be stable. If this conclusion is correct, it seems to be the first
  example of a stable \wh\ model in \GR.

\section{The static model and perturbations}
\label{br-s2}

  The background static metric (\ref{br-ds0}) is well known to be a solution to
  the Einstein equations with a source in the form of a massless phantom
  scalar field [12, 13]. The same metric is also a
  solution to the Einstein equations with a composite source consisting of
  neutral phantom dust with the energy density $\eps$ and a source-free
  electromagnetic field [20], so that the stress-energy
  tensor (SET) is\footnote
    {We are using the signature $(+ - - -)$ and the units where
    $c=1$ (the speed of light) and $G=1$ (the gravitational constant).}
\bearr
    T\mN = \frac{q^2}{8\pi r^4} \diag (1, 1, -1, -1)
        + \eps \diag (1, 0, 0, 0),
\nnn                                                     \label{br-SET-0}
    \eps = - \frac{2q^2}{8\pi r^4}.
\ear
  Here $q$ is the magnetic charge, corresponding to the electromagnetic
  field tensor components $F_{\theta\varphi} = - F_{\varphi\theta} =
  q\sin\theta$ and all other $F\mn =0$ (or, alternatively, an electric
  charge, such that $F_{tx} = -F_{xt} = q/r^2$, with all other $F\mn = 0$).
  The electric or magnetic field exists in the \wh\ space-time without
  sources, the charge $q$ characterizing the density of radial lines of
  force threading the \wh\ throat and extending from one spatial infinity
  ($x = +\infty$) to the other ($x = -\infty$).

  Considering small perturbations of the above configuration, we can
  restrict ourselves to axially symmetric perturbations, independent of
  the azimuthal angle $\varphi$, because the possible $\varphi$ dependence
  of the form $\e^{im\varphi}$ does not affect the perturbation equations.
  (A similar phenomenon is well known for the equations of quantum
  mechanics, and for perturbations of gravitating systems it has been
  discussed, in particular, by Chandrasekhar [19].) So we can
  use the perturbed metric in the general axially symmetric form
\bearr                                                       \label{br-ds1}
    ds^2 = \e^{2\nu} dt^2 - \e^{2\mu_x} dx^2 - \e^{2\mu_\theta}\d\theta^2
\nnn \cm
       - \e^{2\psi}(d\varphi -\sigma dt - q_x dx -q_\theta d\theta)^2,
\ear
  where $\nu,\ \mu_x,\ \mu_\theta,\ \psi,\ q_x,\ q_\theta$ are functions
  of $t,\ x,\ \theta$. The background static configuration is characterized
  by
\bearr
    \e^{2\nu} = \e^{2\mu_x} =1, \qquad \e^{2\mu_\theta}= r^2 = x^2 + q^2,
\nnn
        \e^{2\psi} = r^2 \sin\theta,\qquad \sigma = q_x = q_\theta =0.
\ear

  The perturbed electromagnetic field tensor may contain any small additions
  $\delta F\mn (t,x,\theta)$ to the background $F\mn$ described above. As to
  the matter content, we assume it in the form of a perfect fluid with the
  SET in its standard form
\beq
  	T\mn = (\eps + p)u_\mu u_\nu - p g\mn.
\eeq
  In the background configuration, $\eps$ is given in (\ref{br-SET-0}), and
  $p=0$. For the perturbed configuration we put, following [18],
\bearr
    8\pi\eps = - \frac{2q^2}{(q^2+x^2)^2} + f(t,x,\theta), \label{br-eps1}
\yyy
    8\pi p = h(x) f(t,x,\theta).             \label{br-p1}
\ear
  Thus $f$ characterizes the density perturbation of matter, and $h(x)$
  determines its equation of state.

  The 4-velocity has the form $u_\mu = (1,0,0,0) = u^\mu$ in the
  background, but it may acquire small spatial components $u_i$ in the
  perturbed configuration.

  Small perturbations of the static background split into two classes,
  polar and axial perturbations, depending on their symmetry with respect to
  the reflection $\varphi \mapsto - \varphi$, and these classes can be
  studied independently of each other (see, e.g., [19]).

  Polar perturbation, which are even at $\varphi \mapsto - \varphi$, are
  characterized by nonzero increments $\delta\nu$, $\delta\psi$,
  $\delta\mu_x$, $\delta\mu_\theta$ as well as those of $F_{tx}$,
  $F_{t\theta}$ and $F_{x\theta}$ and nonzero velocity components $u^x$ and
  $u^\theta$.

  Axial perturbations, which are odd at $\varphi \mapsto - \varphi$,
  involve nonzero perturbations $\sigma,\ q_x,\ q_\theta$, $\delta
  F_{\mu\varphi}$ and the velocity component $u^{\varphi}$.

  Each class of perturbations can evidently contain a nonzero increment
  $f(t,x,\theta)$ of matter density.

  In what follows, we will study the monopole modes of polar perturbations,
  which do not violate spherical symmetry, and general axial perturbations
  of the above background. A full consideration of nonspherical polar
  perturbations, which is technically more complicated, is postponed for the
  future.

\section{Stability of the spherical mode}

  In the case of \sph\ perturbations, the only dynamic mode is related to
  a possible motion of matter particles while the gravitational and
  electromagnetic variables are excluded using the Einstein and Maxwell
  equations. This calculation has been performed for the \wh\ under study in
  [18], resulting in the following equation:
\bearr
    \ddot{\eta} - h(x)\eta'' + \biggl[-h' + \frac{2h}{r^2 x}\biggr]\eta'
        + U(x) \eta =0,
\nnn
    U(x)\equiv \frac{12h x^2 + 3h - 3h' x r^2 + 1}{r^4},
\ear
  where the perturbation $\eta(x)$ is defined by the relations
  $\e^{2\mu_x}= r^2\e^{2\eta}$, $\e^{2\psi}= r^2\e^{2\eta} \sin^2 \theta$.
  Here, without loss of generality, we have passed on to dimensionless
  variables by putting formally $q=1$ (lengths are now measured in units
  of the \wh\ throat radius), so that now $r^2 = 1+x^2$. It is assumed
  $h \geq 0,\ h(\pm\infty) =\const >0$, and the potential $U(x)$ then tends
  to zero as $x \to \pm \infty$.

  We separate the variables putting $\eta \sim \e^{i\omega t}$ and write
  down the equation for a separate mode with the frequency $\omega$:
\beq
    h(x)\eta'' + \biggl[h' - \frac{2h}{r^2 x}\biggr]\eta'   \label{br-mode}
                + [\omega^2 - U(x)] \eta =0.
\eeq

  Let us assume $U(x) > 0$ and consider unstable modes with imaginary
  frequencies, $\omega^2 < 0$. Denote $\Omega^2 (x) := -\omega^2 + U(x)
  >0$. Then evidently the physically admissible asymptotics at $x \to \pm
  \infty$ have the form $\eta (x) \propto \e^{-|\omega x|}$.  Suppose
  without loss of generality that at such an asymptotic, as $x\to -\infty$,
  we have $\eta >0$, then at large negative $x$ we inevitably have $\eta'>
  0$, $\eta'' >0$.

  A physically admissible solution $\eta (x)$, whatever be its behavior
  at finite $x$, should return to zero at large $x$, hence it should have
  a maximum at some $x = x_0$, at which $\eta =  \eta_0 > 0$,
  $\eta' =0$, $\eta''<0$.

  It is clear that such a maximum is impossible at $x_0 \ne 0$, where
  the coefficient at $\eta'$ in \eq (\ref{br-mode}) is finite, and the
  equation leads to $\eta'' >0$ at a point where $\eta' =0$.

  At $x_0 =0$ the situation is different due to the singularity $\sim 1/x$
  in the coefficient at $\eta'$. Consider a solution near $x_0 =0$
  as a Taylor series
\beq
    \eta(x) = \eta_0 + \half \eta_2 x^2 + \dots,
\eeq
  corresponding to a minimum if $\eta_2 >0$ and to a maximum if $\eta_2 <0$.

  Assuming $h(0) = h_0 \ne 0$, \eq (\ref{br-mode}) in the order $O(1)$ leads
  to the equality $\eta_2 = - \Omega^2(0) \eta_0/h_0 < 0$, so that a maximum
  is possible, hence an unstable mode of perturbations is also possible.

  Let us assume $h(x) = a x^n + o(x^n)$, $a >0$, $n > 0$ at small $x$,
  then in the senior order of magnitude in $x$ \eq (\ref{br-mode}) gives
\beq                            \label {hx4}
    a(n - 1) \eta_2 x^n - \Omega^2 (0)\eta_0 = 0.
\eeq
  At $n \ne 1$, under the above assumptions $\Omega^2 (0) > 0$, $\eta_0 >0$,
  this equality makes a contradiction, which (provided $U(x) > 0$) proves
  the nonexistence of unstable modes of our system with $\omega^2 < 0$.

  It is not hard to verify, in particular, that the condition $U(x) >0$
  holds at all $x$ if
\beq                                                       \label{br-h_n}
    h(x) = ax^n/r^n, \cm 0 < a < 1,
\eeq
  and $n$ is an even integer from 2 to 14 (if $h(x)$ is even, $U(x)$ is even
  as well).

  One should also consider a possible zero mode, $\omega =0$, at which the
  perturbation can linearly grow with time. Then the physically admissible
  asymptotic behavior of the solution to (\ref{br-mode}) is a decay by a
  power law instead of an exponential,\footnote
    {As $x \to \pm \infty$, under the above assumptions, $h \to h_*$ and
    $U \approx U_*/x^2$, where $0 < h_* \leq 1$ and $U_* >0$. Then the
    substitution $\eta \sim |x|^{-k}$ in (\ref{br-mode}) leads to
    $k(k+1) = U_*/h_*$.}
  $\eta \sim |x|^{-k}$, $k > 0$. However, the further reasoning completely
  repeats that for $\omega^2 < 0$ with the same result.

  We conclude that our background configuration is stable under \sph\
  perturbations for matter with the equation of state involving the function
  (\ref{br-h_n}).

  It should be noted that this result has been obtained without bringing \eq
  (\ref{br-mode}) to the canonical form and without dividing the $x$ axis into
  parts, unlike [18]; moreover, we have included a mode with a nonzero
  perturbation of the throat radius, the most ``dangerous'' one as regards
  the instability.

\section{Axial perturbations}

  As described in Section 2, axial perturbations include (a) perturbations
  of matter density and pressure as well as the velocity directed along the
  azimuthal angle $\varphi$; (b) perturbations of the metric (\ref{br-ds1}),
  including nonzero $\sigma,\ q_x,\ q_\theta$, while perturbations of $\nu,\
  \psi,\ \mu_x$, $\mu_\theta$ are zero; (c) perturbations of the
  electromagnetic field $F_{\varphi\mu}$. It is easy to see that a small
  coordinate transformation $\varphi \to \varphi +
  \delta\varphi(t,x,\theta)$ makes it possible to turn to zero any small
  velocity field $v^\varphi = d\varphi/dt$, directed along $\varphi$.
  Therefore, without loss of generality, we can assume that the fluid is in
  its comoving reference frame, and the velocity field has the form
\beq
      u^\mu = (e^{-\nu}, 0, 0, 0); \ \ \
      u_\mu = (e^{\nu}, 0, 0, \sigma e^{\psi-\nu}).
\eeq

  Consider the fluid equations of motion in terms of its SET $T\mN$:
  $\nabla_\nu T\mN =0$. It is sufficient for us to consider the generalized
  continuity equation $u^\mu \nabla_\nu T\mN =0$, which now reads
\beq
      [(\rho + p) e^{\psi+\mu_x+\mu_\theta}]\dot{}
		 = e^{\psi+\mu_x+\mu_\theta} \dot p.
\eeq

  Since the quantities $\psi,\ \mu_x,\ \mu_\theta$ are not perturbed, this
  equation leads to the equality $(\rho +p)\dot{} = \dot p$, that is,
  $\dot\rho =0$, so that the density (hence also the pressure) have only
  static perturbations, which can be neglected since we are only
  interested in dynamic perturbation modes. Thus matter is not perturbed,
  and we can restrict ourselves to perturbations of free (though mutually
  interacting) electromagnetic and gravitational fields. Such a problem
  has been considered in a general form in [21] for an arbitrary
  \ssph\ background configuration. It has been shown there that,
  after separating the time variable with the factor $\e^{i\omega t}$ and
  the angular variable $\theta$ using the appropriate Gegenbauer functions,
  the dynamic perturbations can be described by two radial functions
  $H_1(x)$ (responsible for electromagnetic perturbations) and $H_2(x)$
  (for the gravitational ones), which obey the equations [21]
\bearr                          \label{br-eq-H1}
    \frac{d^2 H_1}{dx^2}+ \omega^2 H_1
        = \frac{\Delta}{r^4} \left(L^2 + 2
         + \frac{4 q^2}{R^2 }\right) H_1
\nnn \inch
         + \frac{2 q L\Delta}{r^5} H_2,
\\ \lal                         \label{br-eq-H2}
    \frac{d^2 H_2}{dx^2} + \omega^2 H_2 = \left(-\frac{r_{xx}}{r} +
    \frac{2 r_x^2}{r^2} + L^2 \frac{\Delta}{r^4}\right) H_2
\nnn \inch
         + \frac{2 q L \Delta}{r^5} H_1,
\ear
  where $r^2 = e^{2\mu_\theta}$, $\Delta = r^2 e^{2\nu}$,
  $L^2 = (\ell -1)(\ell +2)$, $r_x = \d r/\d x$, $r_{xx} = \d^2 r/\d x^2$,
  $\ell$ is the multipolarity order, and $x$ is the ``tortoise'' radial
  coordinate defined by the condition $g_{tt} = -g_{xx}$.

  In our case of the metric (\ref{br-ds0}), the coordinate $x$ satisfies this
  condition, and substitution of this metric into \eqs (\ref{br-eq-H1}) and
  (\ref{br-eq-H2}) leads to the coupled oscillator equations
\bearr                      \label{br-eq-H}
     	H''_1 + \omega^2 H_1 = V_{11} H_1 + V_{12} H_2,
\nnn
        H''_2 + \omega^2 H_2 = V_{21} H_1 + V_{22} H_2,
\ear
  where
\bearr                      \label{br-V_ab}
    V_{11} = \frac{1}{r^4} \Big[ r^2(L^2 + 2) + 4 q^2\Big],
\nnn
    V_{22} = \fracd{1}{r^4} \Big[ x^2 (L^2 + 2) + q^2 (L^2 - 1) \Big],
\nnn
    V_{12} = V_{21} = \fracd{2 q L}{r^3}.
\ear

  For these equations to admit decoupling, it is necessary and sufficient
  that the matrix $(V_{ab})$ have an eigenvector independent of $x$ (see,
  e.g., [22]). Such a decoupling made it possible to obtain two separate
  wave equations for perturbations of the \RN\ solution [19], but in our
  case this method does not work since the eigenvectors of the matrix
  $V_{ab}$ can be written as
\beq
    \Big\{ 7 q^2 \pm \sqrt{49 q^4 + 16 L^2 q^2 r^2}, \ 4 L q r \Big\}.
\eeq
  and none of them is a multiple of a constant vector, except for the case
  $L=0$ (that is, $\ell =1$) where $V_{12} = V_{21} =0$ and the equations
  are already decoupled.

  If the equations are not decoupled, a sufficient condition for $\omega^2>0$
  under zero boundary conditions at both infinities is that the matrix
  $V_{ab}$ is nonnegative-definite at each $x$, which for a $2 \times 2$
  matrix reduces to the requirements that its trace and determinant are
  nonnegative (see the Appendix). One can directly verify that this is
  indeed the case for $\ell \geq 2$, hence all such modes are stable.

  At $\ell =1$, $V_{12} = V_{21} =0$, and (\ref{br-eq-H}) are two separate
  equations, the first one with the manifestly positive potential $V_{11}$
  guaranteeing stability of the corresponding mode, and the other one with
\[
    V_{22} = (2x^2 -  q^2)/r^4,
\]
  containing a potential well near $x=0$. This potential was considered by
  Armendaris-Picon in [1], and it was shown that the ground state of this
  problem with zero boundary conditions ($H_2 (\pm \infty) =0$) corresponds
  to $\omega =0$. It leads to $\ddot H_2 =0$, hence $H_2$ can linearly
  grow with time, indicating an instability of the background configuration.
  It has been shown, however, in [1] (where perturbations of the same metric
  (\ref{br-ds0}) were considered, though in the presence of a scalar field
  but without an electromagnetic one), that the axial gravitational
  perturbations with $\ell =1$ are actually a pure gauge and can be
  annihilated by a suitable coordinate transformation. It is quite a natural
  observation from a physical viewpoint since the dynamic (wave) degrees of
  freedom of the gravitational field of a compact source begin with
  quadrupole modes ($\ell =2$) while dipole perturbations should be
  stationary. This result applies to our system because the gravitational
  degree of freedom is decoupled from the electromagnetic one in the dipole
  mode.

  We conclude that our \wh\ model is linearly stable under all axial
  perturbations.

\section{Conclusion}
\label{br-s4}

  We have shown that, in \GR, it is possible to construct a static
  traversable Lorentzian \wh\ which is stable under all spherical and
  axial perturbations. The latter turn out to be independent of the equation
  of state assumed for the matter supporting the \wh\ (specifically, on the
  particular choice of $h(x)$).

  The experience of stability studies for different \ssph\ configurations
  shows that nonspherical polar perturbations behave qualitatively in the
  same way as their axial counterparts because, for both kinds of
  perturbations, the wave equations with nonzero multipolarities $\ell$
  possess effective potentials with positive ``centrifugal'' terms [1, 19].
%% \cite{Chand, Armendariz-Picon2002}.
  Therefore if a model is really unstable, the instability will most
  probably manifest itself in spherical modes where $\ell =0$.

  Nevertheless, to complete the stability study of our model, nonspherical
  polar perturbations should be studied, and we are planning to consider
  them in the near future. The most probable result of such a study must
  show that the present model is (to our knowledge, at least for 
  distributed systems as opposed to thin-shell \whs)
  the first example of a stable \wh\ model in \GR.

\subsection*{Appendix. Coupled wave equations: a sufficient condition for
    stability}

\def\vy {{\vec y}\,{}}
\def\vz {{\vec z}\,{}}
\def\theequation{A.\arabic{equation}}
\sequ{0}

  We present this elementary and well-known proof for completeness.
  Consider a set of coupled equations of the form
\beq                                                       \label{br-eq-vy}
       \vy'' + \omega^2 \vy = V(x) \vy, \cm     x \in \R,
\eeq
  where $\omega = \const$, $V = (V_{ab}(x))$ is an $n \times n$ matrix with
  $x$-dependent elements, and $\vy = (y_1(x), y_2(x)$, $\ldots, y_n(x))$ is a
  column of $n$ unknown functions of $x$, which are assumed to be
  square-integrable, so that, in particular, $y_a \to 0$ as $x \to \pm
  \infty$. The prime denotes $d/dx$.

  Consider $\vy$ as a vector in $n$-dimensional Euclidean space with
  the usual scalar product. Let us scalarly multiply \eq (\ref{br-eq-vy}) by
  $\vy$ from the left and integrate over $\R$ to obtain (omitting the limits
  near the integral sign)
\bear
      \int \vy \vy'' dx + \omega^2 \int \vy^2 dx = \int \vy\, (V \vy) dx.
\ear
  The first integral can be rewritten as
\[
      \int (\vy \vy')'\, dx - \int \vy'{}^2 dx = - \int \vy'{}^2 dx,
\]
  since the first term here is directly integrated and vanishes due to
  the boundary condition. As a result, we obtain the following expression
  for $\omega^2$:
\beq
    \omega^2 = \frac{1}{\int \vy^2 dx }\
           \biggl[ \int \vy'{}^2 dx + \int \vy\, (V\vy) dx\biggr].
\eeq
  We assume that the solution $\vy$ is nontrivial, hence both the
  denominator and the first term in the square brackets are nonzero.
  Therefore a sufficient condition for $\omega^2 > 0$ is that
\beq
       \vy\,(V\vy) \equiv V_{ab}(x) y_a(x) y_b(x) \geq 0
\eeq
  at all values of $x$, or, in other words, that this quadratic form is
  nonnegative-definite at all $x$.

  For $n=1$ this reduces to $V \geq 0$, the well-known sufficient condition
  of only positive energy states in a one-dimensional quantum-mechanical
  system.

  For $n=2$ this requirement leads to two conditions: the trace
  $V_{11} + V_{22} \geq 0$ and $\det (V_{ab}) \geq 0$.

\subsection*{Acknowledgments}
\label{br-s5}

  The authors are thankful to the participants of seminars at the Astro
  Space Center of Lebedev Physical Institute of RAN and at the Sternberg
  Astrophysical Institute (Moscow) as well as to Roman Konoplya, Valentina
  Kolybasova and Milena Skvortsova for helpful discussions and remarks.

  The work has been supported in part by RFBR Projects 13-02-00757-a,
  12-02-00276-a , 11-02-00244-a , 13-02-00757-a, 11-02-12168-ofi-m-2011,
  Scientific School 2915.2012.2 (Formation of the Large-Scale
  Structure of the Universe and Cosmological Processes), the Program
  ``Non-stationary Phenomena in Objects of the Universe 2012'', and
  the Program ``Scientific and Pedagogical Personnel of Innovative
  Russia 2009--2013', Federal Goal-Oriented Program 16.740.11.0460.

\end{document}